\documentclass[aps,prl,floatfix,amsmath,amssymb,nofootinbib,twocolumn]{revtex4-1}
\usepackage{graphicx,color,dcolumn,booktabs,bm,adjustbox,esint,upgreek}
\usepackage{longtable,comment,braket}
\usepackage{amsfonts,mathrsfs}
\usepackage{times}
\usepackage{overpic}
\usepackage{tipa}
\usepackage{indentfirst}
\usepackage{feynmf}   
\usepackage{slashed}  
\usepackage{cases}
\usepackage{epstopdf}
\usepackage{psfrag}
\usepackage{subfigure}
\usepackage{capt-of}
\usepackage{dcolumn,kantlipsum}
\usepackage{color}
\usepackage{multirow}
\usepackage{ulem}
\usepackage[colorlinks, citecolor=blue,anchorcolor=red,menucolor=red,linkcolor=blue,filecolor=red,runcolor=red,urlcolor=blue,frenchlinks=red]{hyperref}
\allowdisplaybreaks[4]

\begin{document}
\title{Decoding the nature of $Z_{cs}(3985)$ and establishing the spectrum of charged heavy quarkoniumlike states in chiral effective field theory}
\author{Bo Wang$^{1,2}$}
\author{Lu Meng$^{3}$}\email{lu.meng@rub.de}
\author{Shi-Lin Zhu$^{1,2}$}\email{zhusl@pku.edu.cn}
\affiliation{
$^1$Center of High Energy Physics, Peking University, Beijing 100871, China\\
$^2$ School of Physics and State Key Laboratory of Nuclear Physics and Technology, Peking University, Beijing 100871, China\\
$^3$Ruhr University Bochum, Faculty of Physics and Astronomy,
Institute for Theoretical Physics II, D-44870 Bochum, Germany}

\begin{abstract}
We study the newly observed charmoniumlike state $Z_{cs}(3985)$ in
the framework of chiral effective field theory. The interaction
kernel of the $\bar{D}_sD^\ast/\bar{D}^\ast_sD$ system is calculated
up to the next-to-leading order with the explicit chiral dynamics.
With the fitted parameters extracted from the $Z_c(3900)$ data as
inputs, the mass, width and event distributions of the
$Z_{cs}(3985)$ are very consistent with the experimental
measurements. Our studies strongly support the $Z_{cs}(3985)$ as the
partner of the $Z_c(3900)$ in the SU(3)$_f$ symmetry and the
$\bar{D}_sD^\ast/\bar{D}^\ast_sD$ molecular resonance with the same
dynamical origin as the other charged heavy quarkoniumlike states.
We precisely predict the resonance parameters of the unobserved
states in $\bar{D}_s^\ast D^\ast$, $B^\ast_s
\bar{B}/B_s\bar{B}^\ast$ and $B_s^\ast\bar{B}^\ast$ systems, and
establish a complete spectrum of the charged charmoniumlike and
bottomoniumlike states with the $I(J^P)$ quantum numbers $1(1^+)$
and $\frac{1}{2}(1^+)$, respectively.
\end{abstract}

\maketitle

Very recently, the BESIII Collaboration observed a new charged
charmoniumlike state $Z_{cs}(3985)$ in the $K^+$ recoil-mass
spectrum from the process $e^+e^-\to
K^+(D_s^-D^{\ast0}+D_s^{\ast-}D^0)$ at the center-of-mass energy
$\sqrt{s}=4.681$ GeV~\cite{Ablikim:2020hsk}. Its mass and width were
measured to be
$M_{Z_{cs}}^{\text{pole}}=(3982.5_{-2.6}^{+1.8}\pm2.1)$ MeV and
$\Gamma_{Z_{cs}}^{\text{pole}}=(12.8_{-4.4}^{+5.3}\pm3.0)$ MeV,
respectively. The minimal quark component in $Z_{cs}(3985)^{-}$
should be $c\bar{c}s\bar{u}$ rather than the pure $c\bar{c}$ since
it is a charged particle with strangeness. The mass of
$Z_{cs}(3985)$ is about $100$ MeV larger than that of the
$Z_c(3900)$, which is the typical mass difference between the
$D_s^{(*)}$ and $D^{(*)}$ mesons~\cite{Zyla:2020zbs}. Another
salient feature of $Z_{cs}(3985)$ is the closeness to the
$\bar{D}_sD^\ast/\bar{D}_s^\ast D$ threshold. It is proposed in
Ref.~\cite{Meng:2020ihj} that the newly observed $Z_{cs}(3985)^-$ is
the $U$-spin partner of $Z_c(3900)^-$ under the SU(3)$_f$ symmetry.
The $Z_{cs}(3985)$ has been intensively studied within a very short
time~\cite{Yang:2020nrt,Wang:2020kej,Wan:2020oxt,Du:2020vwb,Chen:2020yvq,Cao:2020cfx,Sun:2020hjw}.

There are large similarities among
$Z_{cs}(3985)$~\cite{Ablikim:2020hsk}, 
$Z_c(3900)$~\cite{Ablikim:2013xfr},
$Z_c(4020)$~\cite{Ablikim:2015vvn}, $Z_b(10610)$ and
$Z_b(10650)$~\cite{Garmash:2015rfd} (we will denote these states as
$Z_{cs}$, $Z_c$, $Z_c^\prime$, $Z_b$ and $Z_b^\prime$, respectively
in the following context for simplicity). They all lie few MeVs
above the corresponding $\bar{D}_sD^\ast$, $D\bar{D}^\ast$,
$D^{\ast}\bar{D}^\ast$, $B\bar{B}^{\ast}$, and
$B^\ast\bar{B}^{\ast}$ thresholds, respectively. They dominantly
decay into the open charm/bottom channels~\cite{Brambilla:2019esw}.
In our recent work~\cite{Wang:2020dko}, we studied the interactions
of the isovector $D^{(\ast)}\bar{D}^{(\ast)}$ and
$B^{(\ast)}\bar{B}^{(\ast)}$ systems with the chiral effective field
theory ($\chi$EFT) up to the next-to-leading order. We find the
invariant mass spectra of the open charm/bottom channels can be
described well and the peaks originate from the poles in the
unphysical Riemann sheet. In other words, the previously observed
$Z_Q^{(\prime)}$ states can be well identified as the dynamically
generated molecular resonances from the $D^{(\ast)}\bar{D}^{(\ast)}$
and $B^{(\ast)}\bar{B}^{(\ast)}$ interactions. The large similarity
between the $Z_{cs}$ and $Z_Q^{(\prime)}$ stimulates us to wonder
whether the newly observed $Z_{cs}$ has the same origin. This Letter
is devoted to answering this question.

The discoveries of more and more near-threshold exotic
states~\cite{Chen:2016qju,Guo:2017jvc,Liu:2019zoy,Lebed:2016hpi,Esposito:2016noz,Brambilla:2019esw}
indicate some common features of QED and QCD. For the very
near-threshold bound states and resonances, physical observables are
insensitive to the details of the interaction, which yields
universality in both hadronic and atomic
sectors~\cite{Braaten:2004rn}. Meanwhile, the hadronic molecules
arise from the residual strong interactions between two color
singlet objects, which is analogous to molecules bound by the
residual interaction of QED. The separable scales in these
near-threshold states lead to a feasible approach to improvable
expansion, which is the basic idea of effective field theory.

The $\chi$EFT is generally accepted as the modern theory of nuclear
forces~\cite{Bernard:1995dp,Epelbaum:2008ga,Epelbaum:2019kcf,Machleidt:2011zz,Meissner:2015wva,Hammer:2019poc,RodriguezEntem:2020jgp},
which is built upon two pioneer works of
Weinberg~\cite{Weinberg:1990rz,Weinberg:1991um}, and has been
successfully applied to describe the low energy $NN$ scatterings,
light and medium
nuclei~\cite{Epelbaum:2008ga,Machleidt:2011zz,Epelbaum:2019kcf}.
Recently, we generalized the framework of $\chi$EFT to the systems
with heavy quarks and reproduced the hidden-charm pentaquarks
successfully ~\cite{Wang:2019ato,Meng:2019ilv}. Within
$\chi$EFT, we predicted the existence of strange hidden charm
molecular pentaquarks $P_{cs}$ in the isoscalar $\Xi_c\bar{D}^\ast$
system~\cite{Wang:2019nvm}. Our prediction was confirmed by the new
measurement of LHCb at the $J/\psi\Lambda$ final
state~\cite{Wang:2020nvm}. Therefore, it is reliable to utilize the
$\chi$EFT to depict the chiral dynamics inside the
$\bar{D}_sD^\ast/\bar{D}_s^\ast D$ systems, likewise. The
investigation could be extended further to $Z_{cs}^\prime$ in the
$\bar{D}_s^\ast D^\ast$ system, as well as their twin partners in
the $B^\ast_s \bar{B}/B_s\bar{B}^\ast$ and $B_s^\ast\bar{B}^\ast$
systems under the heavy quark symmetry. Searching for these states
would help us to assemble the jigsaw puzzles of dynamical details of
the {\it hadronic molecular physics}.

Since these states are produced near the corresponding thresholds,
the interaction potential of a $\mathtt{VP}$ system ($\mathtt{V}$
and $\mathtt{P}$ denote the vector and  pseudoscalar mesons,
respectively) with the fixed isospin can be parameterized in the
nonrelativistic form,
\begin{eqnarray}
\mathcal{V}&=&\sum_{i=1}^6
V_i(\bm{p}^\prime,\bm{p})\mathcal{O}_i(\bm{p}^\prime,\bm{p},\bm{\varepsilon},\bm{\varepsilon}^\dagger),\label{VOperator1}
\end{eqnarray}
where $\bm p$ and $\bm p^\prime$ represent the momenta of initial
and final states in the center of mass system (c.m.s), respectively.
$\bm{\varepsilon}$ and $\bm{\varepsilon}^{\dagger}$ denote the
polarization vectors of the initial and final vector mesons,
respectively. $V_i$ are the scalar functions to be
obtained from the chiral Lagrangians, while $\mathcal{O}_i$ are six
pertinent operators:
\begin{eqnarray}
\mathcal{O}_1&=&\bm{\varepsilon}^\dagger\cdot\bm{\varepsilon},\quad\mathcal{O}_2=(\bm{\varepsilon}^\dagger\times\bm{\varepsilon})\cdot(\bm q\times\bm k),\nonumber\\
\mathcal{O}_3&=&(\bm q\cdot\bm{\varepsilon}^\dagger)(\bm q\cdot\bm{\varepsilon}),\quad\mathcal{O}_4=(\bm k\cdot\bm{\varepsilon}^\dagger)(\bm k\cdot\bm{\varepsilon}),\nonumber\\
\mathcal{O}_5&=&(\bm q\times\bm{\varepsilon}^\dagger)\cdot(\bm
q\times\bm{\varepsilon}),\quad\mathcal{O}_6=(\bm
k\times\bm{\varepsilon}^\dagger)\cdot(\bm k\times\bm{\varepsilon}),
\end{eqnarray}
with $\bm q=\bm p^\prime-\bm p$ the transferred momentum and $\bm
k=(\bm p^\prime+\bm p)/2$ the average momentum.

Within the framework of $\chi$EFT, the potential up to the NLO in
the paired $\bar{D}_sD^\ast/\bar{D}_s^\ast D$ system can be
classified as the contact interaction, one-eta-exchange (OEE) and
two-kaon-exchange (TKE) contributions. The contact potential
$\mathcal{V}_{\text{ct}}$ is parameterized order by order in power
series of $\bm q$ and $\bm k$ as,
\begin{eqnarray}\label{Vct}
\mathcal{V}_{\text{ct}}&=&(C_0+C_1\bm q^2+C_2\bm
k^2)\mathcal{O}_1+\sum_{i=2}^{6}C_{i+1}\mathcal{O}_i+\dots,
\end{eqnarray}
where $C_i~(i=0,\dots,7)$ are the unknown low energy constants
(LECs), and the ellipsis denotes the higher order terms.

The effective potentials arising from the OEE and TKE contributions
can be extracted from the LO chiral Lagrangians,
\begin{eqnarray}\label{LagLO}
\mathcal{L}&=&i\langle\mathcal{H}v\cdot\mathcal{D}\bar{\mathcal{H}}\rangle+g\langle\mathcal{H}\gamma^\mu\gamma_5u_\mu\bar{\mathcal{H}}\rangle\nonumber\\
&&-i\langle
\bar{\tilde{\mathcal{H}}}v\cdot\mathcal{D}\tilde{\mathcal{H}}\rangle+g\langle
\bar{\tilde{\mathcal{H}}}\gamma^\mu\gamma_5u_\mu
\tilde{\mathcal{H}}\rangle,
\end{eqnarray}
where the covariant derivative
$\mathcal{D}_\mu=\partial_\mu+\Gamma_\mu$. The $\mathcal{H}$ and
$\tilde{\mathcal{H}}$ denote the superfields of the charmed and
anticharmed mesons, respectively. One can consult
Refs.~\cite{Wang:2018atz,Wang:2019nvm,Wise:1992hn} for their
expressions. The axial coupling $g\simeq0.57$ is extracted from the
partial decay width of $D^{\ast+}\to D^0\pi^+$~\cite{Zyla:2020zbs}.
The chiral connection $\Gamma_\mu$ and axial-vector current $u_\mu$
are formulated as
\begin{eqnarray}\label{GammaU}
\Gamma_\mu&\equiv&\frac{1}{2}\left[\xi^\dag,\partial_\mu
\xi\right],\qquad u_\mu\equiv\frac{i}{2}\left\{\xi^\dag,\partial_\mu
\xi\right\},
\end{eqnarray}
where $\xi^2=U=\exp\left(i\varphi/f_\varphi\right)$, with $\varphi$
the normally used matrix form of the light Goldstone
octet~\cite{Wang:2019nvm}, and the decay constants
$f_K=113$ and $f_\eta=116$ MeV, respectively.

Now, the quantum number $I^G(J^{PC})=1^+(1^{+-})$ for the $Z_c$
state is favored~\cite{Zyla:2020zbs} ($C$ parity for the neutral
one). The $J^P$ quantum number of the $Z_{cs}$ is undetermined, but
$I(J^P)=\frac{1}{2}(1^+)$ is presumably used in most
works~\cite{Ablikim:2020hsk,Meng:2020ihj,Yang:2020nrt,Wan:2020oxt}.
Under this assumption, the flavor wave function of the $Z_{cs}$
reads~\cite{Meng:2020ihj}
\begin{eqnarray}\label{Flav}
|Z_{cs}^-\rangle=\frac{1}{\sqrt{2}}(|D^{\ast-}_sD^0\rangle+|D_s^-D^{\ast0}\rangle).
\end{eqnarray}
One can easily get the OEE potential,
\begin{eqnarray}
\mathcal{V}_{\text{OEE}}&=&-\frac{g^2}{6f_\eta^2}\frac{\mathcal{O}_3}{\bm
q^2+m_\eta^2},\label{VOEE}
\end{eqnarray}
where $m_\eta$ is the $\eta$ meson mass, and $\bm
q^2=p^2+p^{\prime2}-2pp^\prime\cos\vartheta$ ($p=|\bm p|$,
$p^\prime=|\bm p^\prime|$, and $\vartheta$ is the scattering angle
in the c.m.s of $\mathtt{VP}$).

The TKE potential from the loop diagrams (see
Ref.~\cite{Wang:2020dko} for the involved loop diagrams, and
Refs.~\cite{Wang:2018atz,Wang:2019ato} for the calculation details)
can be simplified into a compact form in the heavy quark limit and
SU(3)$_f$ limit,
\begin{eqnarray}
\mathcal{V}_{\text{TKE}}&=&V_{1}\mathcal{O}_1,\label{VTPE}
\end{eqnarray}
with
\begin{eqnarray}\label{VTPEform}
V_{1}&=&-\frac{24(4g^2+1)m_K^2+(38g^2+5)\bm q^2}{2304\pi^2f_K^4}\nonumber\\
&&+\frac{6(6g^2+1)m_K^2+(10g^2+1)\bm q^2}{768\pi^2f_K^4}\ln\frac{m_K^2}{(4\pi f_K)^2}\nonumber\\
&&+\frac{4(4g^2+1)m_K^2+(10g^2+1)\bm q^2}{384\pi^2f_K^4y}\varpi\arctan\frac{y}{\varpi},
\end{eqnarray}
where $\varpi=\sqrt{\bm q^2+4m_K^2}$, and
$y=\sqrt{2pp^\prime\cos\vartheta-p^2-p^{\prime2}}$. This result is
obtained with the dimensional regularization, and the divergence is
absorbed by the unrenormalized LECs introduced in Eq.~\eqref{Vct}.

\begin{figure}[!hptb]
\begin{centering}
    \scalebox{1.0}{\includegraphics[width=0.95\linewidth]{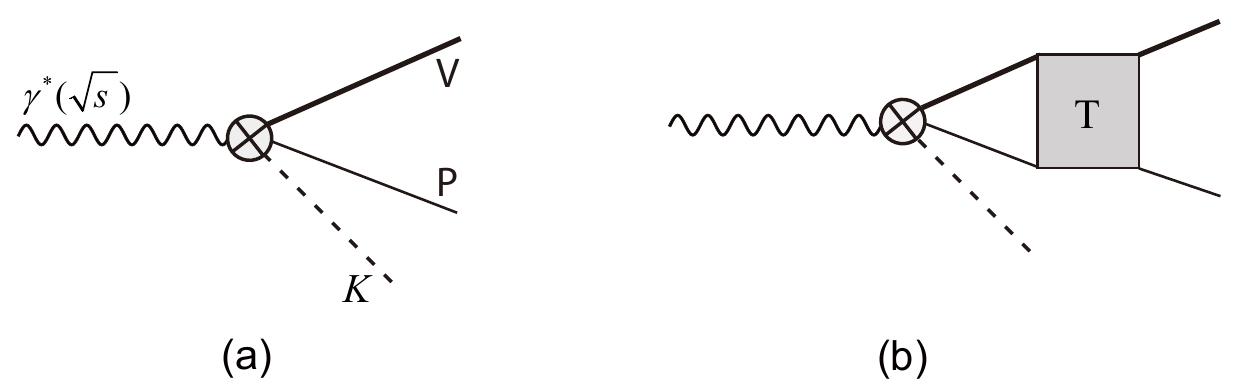}}
    \caption{Diagrams (a) and (b) describe the direct production and rescattering contribution, respectively. The wiggly, thick, thin and dashed lines denote the virtual photon, vector meson, pseudoscalar meson and kaon, respectively. The gray circle with cross stands for the effective $\gamma^\ast\to K\mathtt{VP}$ coupling, while the gray box in diagram (b) signifies the rescattering $T$ matrix of the $\mathtt{VP}$ system.\label{Production}}
\end{centering}
\end{figure}

The $Z_{cs}$ state is observed in the three-body decay of
$e^+e^-\to\gamma^\ast\to K^+(D_s^-D^{\ast0}+D_s^{\ast-}D^0)$. By
fitting the line shape of the $K^+$ recoil-mass spectrum, we can
extract the resonance parameters and pin down the inner structure of
this state. The reaction is illustrated in Fig.~\ref{Production},
where the diagrams \ref{Production}(a) and \ref{Production}(b)
depict the direct production and rescattering effect, respectively.
The rescattering in Fig.~\ref{Production}(b) can generate the
$Z_{cs}$ state dynamically. Additionally, we construct the following
effective Lagrangians to mimic the $\gamma^\ast\to K\mathtt{VP}$
coupling vertex,
\begin{eqnarray}\label{prodvec}
\mathcal{L}_{\gamma^\ast\varphi VP}&=&g_\gamma\mathcal{F}^{\mu\nu}\Big[(\tilde{P}_\mu^\dagger u_\nu P^\dagger-\tilde{P}_\nu^\dagger u_\mu P^\dagger)\nonumber\\
&&-(\tilde{P}^\dagger u_\mu P^\dagger_\nu-\tilde{P}^\dagger u_\nu
P^\dagger_\mu)\Big]+\text{H.c.},
\end{eqnarray}
where $g_\gamma$ denotes the effective coupling constant, and
$\mathcal{F}^{\mu\nu}$ is the field strength tensor of the virtual
photon. $(\tilde{P}_\mu/\tilde{P})$ $P_\mu/P$ denote the
(anti)charmed vector/pseudoscalar meson fields (e.g., see
Ref.~\cite{Wang:2019nvm}). $u_\mu$ is the axial-vector field defined
in Eq.~\eqref{GammaU}.

With the above effective potentials, the $K\mathtt{VP}$ production
amplitude $\mathcal{U}(E,\bm p)$ can be obtained by solving the following
Lippmann-Schwinger equation (LSE),
\begin{eqnarray}\label{LSE}
\mathcal{U}(E,\bm p)&=&\mathcal{M}(E,\bm p)
+\int\frac{d^3\bm q}{(2\pi)^3}\mathcal{V}(E,\bm p,\bm
q)\mathcal{G}(E,\bm q)\mathcal{U}(E,\bm q),\nonumber\\
\end{eqnarray}
where $\mathcal{M}(E,\bm p)$ represents the direct production
amplitude of $\gamma^\ast\to K\mathtt{VP}$ described in
Eq.~\eqref{prodvec}, $E$ is the invariant mass of the $\mathtt{VP}$
pair. The two-body propagator $\mathcal{G}(E,\bm q)$ of the
intermediate state is given as
\begin{eqnarray}\label{GreenF}
\mathcal{G}(E,\bm q)=\frac{2\mu}{\bm p^2-\bm q^2+i\epsilon},\quad
|\bm p|=\sqrt{2\mu(E-m_{\text{th}})},
\end{eqnarray}
with $\mu$ and $m_{\text{th}}$ the reduced mass and threshold of the
$\mathtt{VP}$ system, respectively. In the calculation, we introduce the 
Gaussian form factor $\exp(-p^{\prime2}/\Lambda^2-p^{2}/\Lambda^2)$
(where $\Lambda$ is the cutoff parameter) to avoid heavily involving
the ultraviolet
contributions~\cite{Machleidt:2011zz,RodriguezEntem:2020jgp,Epelbaum:2004fk}.

The LSE of Eq.~\eqref{LSE} is a three dimension integral equation,
which can be reduced to one dimension through the partial wave
decomposition. For example, the effective potentials in
Eqs.~\eqref{Vct} and \eqref{VOEE}-\eqref{VTPEform} can be projected
into the $|\ell sj\rangle$ basis (where $\ell$, $s$ and $j$
represent the orbital angular momentum, total spin and total angular
momentum of the $\mathtt{VP}$ system, respectively)
via~\cite{Golak:2009ri}
\begin{eqnarray}\label{PWD}
\mathcal{V}_{\ell,\ell^\prime}&=&\int d\hat{\bm p}^\prime\int d\hat{\bm p}\sum_{m_{\ell^\prime}=-\ell^\prime}^{\ell^\prime}\langle \ell^\prime,m_{\ell^\prime};s,m_j-m_{\ell^\prime}|j,m_j\rangle\nonumber\\
&&\times\sum_{m_{\ell}=-\ell}^{\ell}\langle \ell,m_{\ell};s,m_j-m_\ell|j,m_j\rangle\mathcal{Y}_{\ell^\prime m_{\ell^\prime}}^\ast(\theta^\prime,\phi^\prime)\nonumber\\
&&\times\mathcal{Y}_{\ell m_\ell}(\theta,\phi)\langle
s,m_j-m_{\ell^\prime}|\mathcal{V}|s,m_j-m_\ell\rangle,
\end{eqnarray}
with $\mathcal{Y}_{\ell m_\ell}$ the spherical harmonics. We have
demonstrated that the $S$-$D$ wave mixing effect is insignificant
for the $Z_Q^{(\prime)}$ states~\cite{Wang:2020dko}, so we only
consider the $S$-wave interaction for the $Z_{cs}$ state in this
Letter. The contact interaction in the $S$-wave projection reads
\begin{eqnarray}
\mathcal{V}_{\text{ct}}&=&\tilde{C}_\text{s}+C_\text{s}(p^2+p^{\prime2}),
\end{eqnarray}
where $\tilde{C}_s$ and $C_s$ are the so-called partial wave LECs.
They are the linear combinations of the LECs introduced in
Eq.~\eqref{Vct}.

The differential decay width for $\gamma^\ast\to K \mathtt{VP}$ can
be expressed in terms of the production amplitude in Eq.~\eqref{LSE}
as
\begin{eqnarray}\label{diffGamma}
\frac{d\Gamma}{dE}=\frac{1}{12(\sqrt{s})^2(2\pi)^3}|\mathcal{U}(E)|^2|\bm
k_1||\bm k_2^\ast|,
\end{eqnarray}
where $\sqrt{s}$ is the c.m.s energy of the $e^+e^-$ collision. $\bm
k_1$ and $\bm k_2^\ast$ are the three momentum of the spectator $K$
in the c.m.s of $e^+e^-$ and the three momentum of $\mathtt{P(V)}$
in the c.m.s of $\mathtt{VP}$, respectively.

\begin{figure}[!hptb]
\begin{centering}
\scalebox{1.0}{\includegraphics[width=0.9\linewidth]{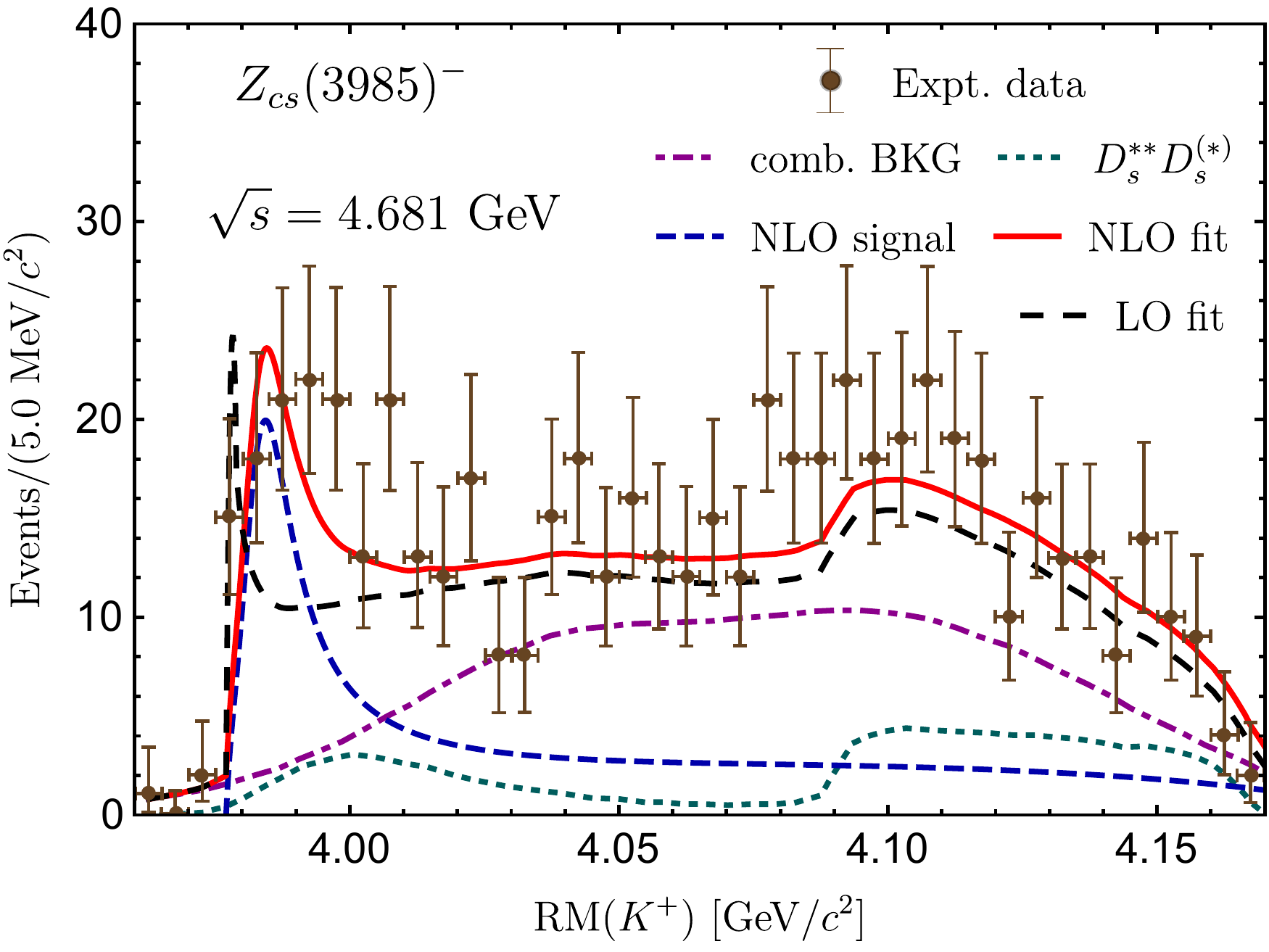}}
\caption{The fit of the $K^+$ recoil-mass spectrum distributions in
the $e^+e^-\to K^+(D_s^-D^{\ast0}+D_s^{\ast-}D^0)$ transitions. The
data with error bars are taken from Ref.~\cite{Ablikim:2020hsk} at
$\sqrt{s}=4.681$ GeV. The red solid, blue dashed, purple dot-dashed,
cyan dotted and black dashed lines denote the NLO fit, signal
contribution with NLO potentials, combinatorial background,
$D_s^{\ast\ast}D_s^{(\ast)}$ contributions (extracted from the
experimental measurements), and LO fit (LO contact plus OEE),
respectively. The line shape of the signal is obtained using the
fitted parameters of $Z_c(3900)$ in Ref.~\cite{Wang:2020dko} as
inputs, i.e., $\tilde{C}_{\text{s}}=3.6$ GeV$^{-2}$,
$C_{\text{s}}=-76.9$ GeV$^{-4}$, and $\Lambda=0.33$ GeV (where the
$\tilde{C}_{\text{s}}$ and $C_{\text{s}}$ are in units of $10^2$,
and the central values in Ref.~\cite{Wang:2020dko} are
used).\label{FitZcs}}
\end{centering}
\end{figure}

We study the $d\Gamma/dE$ distributions and extract the resonance
parameters of the $Z_{cs}$ state. The general procedure is to fit
the $K^+$ recoil-mass spectrum measured by the BESIII
Collaboration~\cite{Ablikim:2020hsk}. We essentially have three free
parameters $\tilde{C}_{\text{s}}$, $C_{\text{s}}$ and $\Lambda$ that
can be varied to match the $K^+$ recoil-mass spectrum. In our
previous work~\cite{Wang:2020dko}, these three parameters are well
fixed by fitting the $D^0D^{\ast-}$ invariant mass distributions of
the $Z_c$ state~\cite{Ablikim:2015swa} (The double $D$ tag technique
is used in this experimental analysis, in which the background
contribution is largely suppressed). When the values of LECs and
cutoff in Ref.~\cite{Wang:2020dko} are fed into the
$D_s^-D^{\ast0}/D_s^{\ast-}D^0$ systems, we find a sharp peak
automatically emerges around $3.98$ GeV in the $D_s^{\ast-}D^0$
invariant mass spectrum. The result is shown in Fig.~\ref{FitZcs},
where the blue dashed line is the production contributions in
Fig.~\ref{Production}. When the other incoherent contributions in
experiments are added up, the total line shape can quantitatively
describe the distributions of experimental events (the red solid
line in Fig.~\ref{FitZcs} with $\chi^2/\text{dof}\simeq0.67$). In
other words, we can describe these two states in an uniform
framework with the same set of parameters, which strongly supports
that the $Z_{cs}$ and $Z_c$ states are partners in SU(3)$_f$
symmetry. We have tried to re-fit the experimental data of
Ref.~\cite{Ablikim:2020hsk}, and find the result is very similar.
The line shape is slightly shifted and the parameters have similar
size but just with a little larger errors (with
$\chi^2/\text{dof}\simeq0.64$). So the outputs are given in terms of
the fitted parameters of the $Z_c$ state in
Ref.~\cite{Wang:2020dko}.

The fit with the LO potentials alone (LO contact terms plus the OEE)
cannot reproduce the experimental data well (the black dashed line
in Fig.~\ref{FitZcs} with $\chi^2/\text{dof}=1.21$), and cannot
describe the event distributions around $3.98$ GeV. Including the
NLO contributions in effective potential gives rise to a resonance
peak, which conforms to the bump structure around $3.98$ GeV in
experiments. The improvement of the fitting indicates that the
$\chi$EFT in the hidden charm sector tends to be convergent.

The peak lies above the $D_s^{\ast-}D^0$ threshold, which
corresponds to a pole of the production $\mathcal{U}$ matrix in the
unphysical Riemann sheet. This can be conducted through analytical
continuation of the Green's function $\mathcal{G}$ defined in
Eq.~\eqref{GreenF},
\begin{eqnarray}
\mathcal{G}^{b}(p+i\epsilon)&\equiv&\mathcal{G}^{a}(p+i\epsilon)-2i\text{Im}\mathcal{G}^{a}(p+i\epsilon),
\end{eqnarray}
where $\mathcal{G}^{a}$ and $\mathcal{G}^{b}$ denote the Green's
function defined in the physical and unphysical Riemann sheets,
respectively. The pole position $m-i\Gamma/2$ reads,
\begin{eqnarray}
(m,\Gamma)&=&\left(3982.4^{+4.8}_{-3.4},
11.8^{+5.5}_{-5.2}\right)~\mathrm{MeV},
\end{eqnarray}
where the errors inherit from those of the fitted parameters in
Ref.~\cite{Wang:2020dko}. The mass and width are highly consistent
with the experimental data~\cite{Ablikim:2020hsk}. Therefore, our
studies strongly support the $Z_c$ and $Z_{cs}$ as the SU(3)$_f$
symmetry parters and the resonances generated from the
$D\bar{D}^\ast/\bar{D}D^\ast$ and $\bar{D}_sD^\ast/\bar{D}_s^\ast D$
interactions, respectively.

In addition, the formations and decay properties of these resonances
can be synchronously interpreted in the molecular configuration
pictures (The compact tetraquarks do not necessarily require their
masses reside very close to the threshold). The near-threshold
production indicates the $\mathtt{V}$ and $\mathtt{P(V)}$ mesons
move very slowly, which renders them have enough time to interact
with each other. A strongly attractive interaction can confine two
particles for infinite time, which corresponds to a stable bound
state. If the attraction is not enough strong but with a barrier to
trap two particles for a finite time, then a resonance with certain
lifetime is generated. In contrast to the bound states, the
resonances naturally decompose into their ingredients at the end of
their lifetime, i.e., the elastic decay modes would contribute
dominantly to the partial decay widths. Yet, the inelastic decays
with final states of a heavy quarkonium and a light meson proceed
via shorter distance interactions (with $r\sim 1/m_D$), which is
generally suppressed and thus reacts with less probability. Thus the
inelastic channels only contribute a small amount of the partial
widths~\cite{Ablikim:2013xfr,Garmash:2015rfd}.
\begin{table*}[htbp]
\centering
\renewcommand{\arraystretch}{1.5}
\caption{The resonance information of $Z_{cs}(3985)$~\cite{Ablikim:2020hsk} and other
predicted states in the $\bar{D}_s^\ast  D^\ast $, $B^\ast_s
\bar{B}/B_s\bar{B}^\ast$ and $B^\ast_s \bar{B}^\ast $ systems. The
superscript `$\dagger$' means this state has been observed, while
the unobserved states are marked with boldface. We define the masses
and widths of the resonances from their pole positions
$E=m-i\Gamma/2$ (with $m$ the mass and $\Gamma$ the width). The
$\Delta m$ represents the distance between the resonance and its
threshold, i.e., $\Delta m=m-m_{\text{th}}$.\label{FitRes}}
\setlength{\tabcolsep}{3.15mm} {
\begin{tabular}{ccccccc}
\hline\hline
Systems&$I(J^P)$&Thresholds [MeV]&Masses [MeV]&Widths [MeV]&$\Delta m$ [MeV]&States\\
\hline
$\frac{1}{\sqrt{2}}[\bar{D}^{\ast}_s D+\bar{D}_sD^{\ast}]$&$\frac{1}{2}(1^+)$&$3977.0$&$3982.5^{+1.8}_{-2.6}\pm2.1$&$12.8^{+5.3}_{-4.4}\pm3.0$&$5.5^{+1.8}_{-2.6}\pm2.1$&$Z_{cs}(3985)^\dagger$\\
$\bar{D}_s^\ast  D^\ast $&$\frac{1}{2}(1^+)$&$4119.1$&$4124.2_{-3.7}^{+5.6}$&$9.8^{+5.2}_{-4.8}$&$5.1_{-3.7}^{+5.6}$&$\bm{Z_{cs}(4125)}$\\
$\frac{1}{\sqrt{2}}[B^\ast_s \bar{B}+B_s\bar{B}^\ast]$&$\frac{1}{2}(1^+)$&$10694.7$&$10701.9^{+3.9}_{-2.7}$&$7.4^{+3.6}_{-4.4}$&$7.2^{+3.9}_{-2.7}$&$\bm{Z_{bs}(10700)}$\\
$B^\ast_s \bar{B}^\ast $&$\frac{1}{2}(1^+)$&$10740.1$&$10747.0^{+4.3}_{-3.1}$&$7.3^{+3.7}_{-4.6}$&$6.9^{+4.3}_{-3.1}$&$\bm{Z_{bs}(10745)}$\\
\hline\hline
\end{tabular}
}
\end{table*}

\begin{figure}
\begin{minipage}[t]{0.49\linewidth}
\centering
\includegraphics[width=\columnwidth]{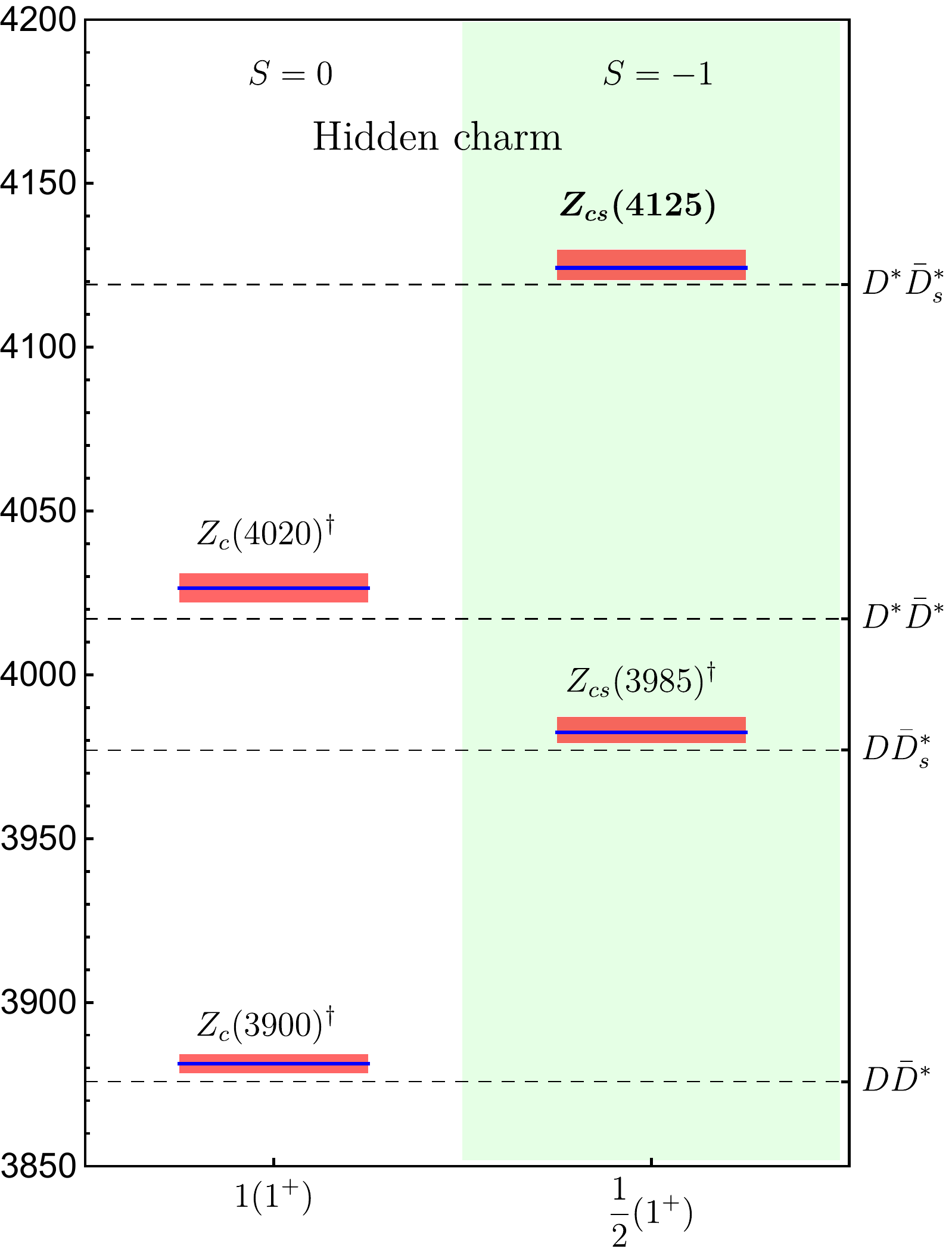}
\end{minipage}%
\begin{minipage}[t]{0.5\linewidth}
\centering
\includegraphics[width=\columnwidth]{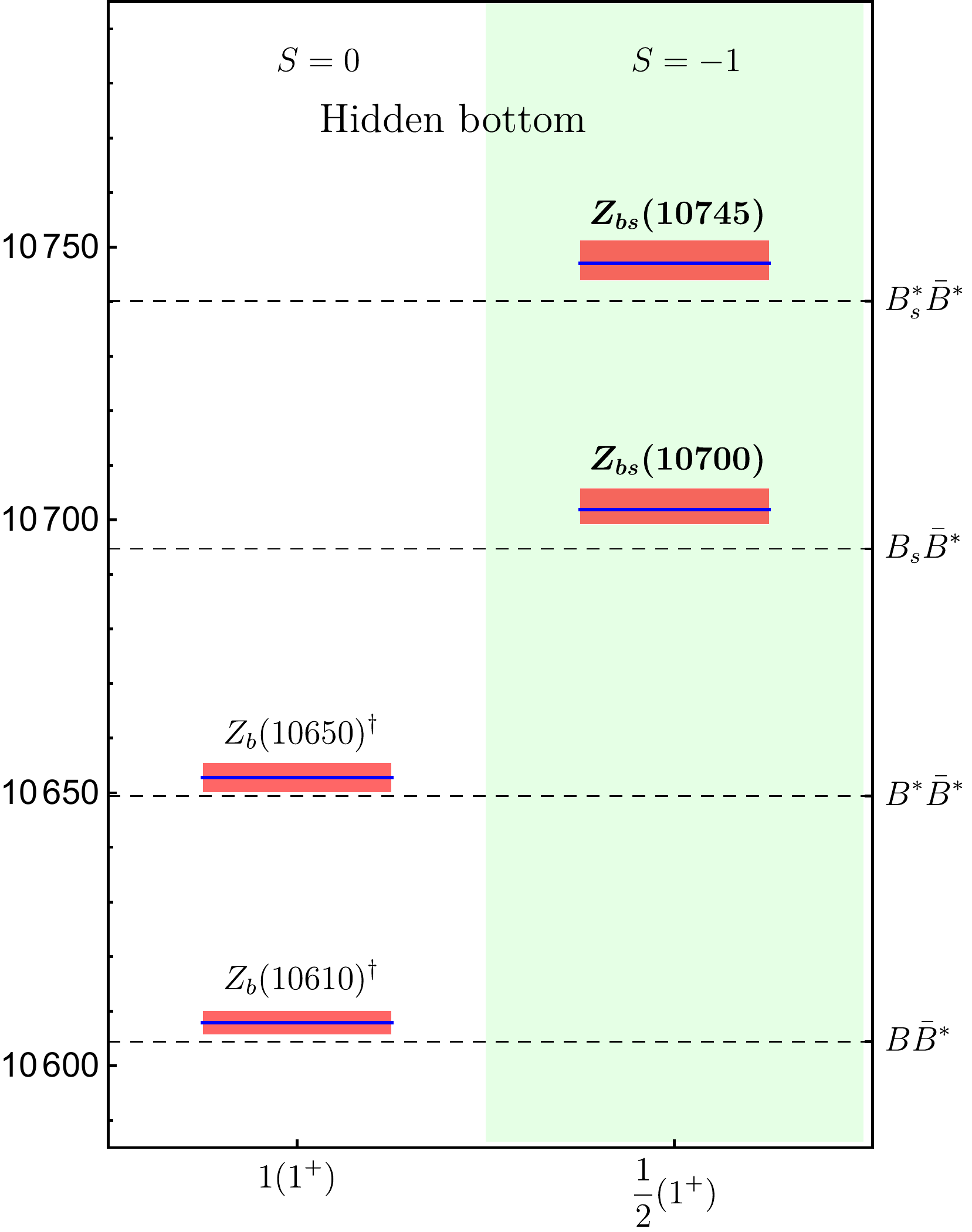}
\end{minipage}
\caption{A complete spectrum of the charged charmoniumlike (left
panel) and bottomoniumlike (right panel) states with strangeness
$S=0$ and $S=-1$, respectively. The blue solid lines and red bands
denote the central values and range of errors of the masses,
respectively. The observed and predicted states are marked with
`$\dagger$' and boldface, respectively.\label{Spectrum}}
\end{figure}

We can adopt the same framework to predict the unobserved states in
the $\bar{D}_s^\ast  D^\ast $ system as well as the $B^\ast_s
\bar{B}/B_s\bar{B}^\ast$ and $B_s^\ast\bar{B}^\ast$ systems in the
hidden bottom sectors. The inputs for these systems come from the
fitted parameters of the $Z_c^\prime$, $Z_b$ and $Z_b^\prime$ states
in Ref.~\cite{Wang:2020dko}, respectively. The predictions are
listed in Table~\ref{FitRes}. We find that there indeed exists a
resonance in the $\bar{D}_s^\ast  D^\ast $ system and two resonances
in the $B^\ast_s \bar{B}/B_s\bar{B}^\ast$ and $B_s^\ast\bar{B}^\ast$
systems, respectively. They lie around $5\sim7$ MeV above the
corresponding thresholds, and their widths coincide with those of
the observed partners. Including the observed $Z_Q^{(\prime)}$
states, we can establish a complete spectrum for the $1(1^+)$ and
$\frac{1}{2}(1^+)$ charged heavy quarkoniumlike states. The spectrum
is vividly illustrated in Fig.~\ref{Spectrum}. These predicted
states could be reconstructed at the corresponding open charm/bottom
channels or the $J/\psi K$ and $\Upsilon(nS)K~(n=1,2)$ final states,
respectively. Hunting for these states would be an intriguing topic
in future experiments.

In summary, we have generalized the framework of $\chi$EFT to decode
the nature of the newly observed exotic $Z_{cs}$ state by
BESIII~\cite{Ablikim:2020hsk}. The proximity to the
$\bar{D}_sD^\ast/\bar{D}_s^\ast D$ threshold and large similarity
with $Z_c$ hint that this unusual state may be a cousin of the $Z_c$
in SU(3)$_f$ family. The interaction kernel of the $\bar{D}_s
D^\ast$ system is calculated up to the NLO, which incorporates the
LO contact terms, OEE and the NLO contact terms and TKE. When the
LECs and cutoff fitted from the $Z_c$ data are fed into the
$Z_{cs}$, iterating the effective potential in LSE automatically
generates a sharp peak near the $\bar{D}_sD^\ast/\bar{D}_s^\ast D$
threshold in the $\bar{D}_s D^\ast$ invariant mass spectrum. The
mass and width from the pole of the production $\mathcal{U}$ matrix
is very consistent with the experimental data, and the distributions
of events can also be well described. Our studies strongly support
that the $Z_{cs}$ and $Z_c$ are partners in SU(3)$_f$ family, and
they have the same dynamical origin. Inspired by the $Z_{cs}$
results, we also predict the resonance parameters of three
unobserved states in the $\bar{D}_s^\ast D^\ast $, $B^\ast_s
\bar{B}/B_s\bar{B}^\ast$ and $B_s^\ast\bar{B}^\ast$ systems. We have
established a complete spectrum of the charged charmoniumlike and
bottomoniumlike states. Looking for these predicted states in
experiments would help us to understand the chiral dynamics, the
manifestations of SU(3)$_f$ and heavy quark symmetries
at the hadron levels, more deeply.\\

\section*{acknowledgements}
This project is supported by the National Natural Science Foundation
of China under Grant 11975033. This work is supported in part by DFG
and NSFC through funds provided to the Sino-German CRC 110
``Symmetries and the Emergence of Structure in QCD" .

\end{document}